\documentclass[a4paper,11pt]{article}
\pdfoutput=1 

\usepackage{jinstpub} 

\title{\boldmath Reconstruction in an imaging calorimeter for HL-LHC}

\author[a,b,1]{A. Di Pilato\note{Corresponding author.},}
\author[c]{Z. Chen,}
\author[d]{F. Pantaleo}
\author[d]{and M. Rovere}

\affiliation[a]{University of Bari,\\Piazza Umberto I, 1, 70121 Bari, Italy}
\affiliation[b]{National Institute for Nuclear Physics (INFN) - Sezione di Bari,\\Via Amendola 173, 70126 Bari, Italy}
\affiliation[c]{Northwestern University,\\633 Clark Street, Evanston, IL 60208, US}
\affiliation[d]{European Organization for Nuclear Research (CERN),\\Esplanade des Particules 1, 1211 Meyrin, Switzerland}

\emailAdd{antonio.dipilato@ba.infn.it}

\abstract{The CMS endcap calorimeter upgrade for the High Luminosity LHC in 2027 uses silicon sensors to achieve radiation tolerance, with the further benefit of a very high readout granularity. Small scintillator tiles with individual SiPM readout are used in regions permitted by the radiation levels. A reconstruction framework is being developed to fully exploit the granularity and other significant features of the detector like precision timing, especially in the high pileup environment of HL-LHC. An iterative clustering framework (TICL) has been put in place, and is being actively developed. The framework takes as input the clusters of energy deposited in individual calorimeter layers delivered by the CLUE algorithm, which has recently been revised and tuned. Mindful of the projected extreme pressure on computing capacity in the HL-LHC era, the algorithms are being designed with modern parallel architectures in mind. Important speedup has recently been obtained for the clustering algorithm by running it on GPUs. Machine learning techniques are being developed and integrated into the reconstruction framework. This paper will describe the approaches being considered and show first results.}

\keywords{high granularity, clustering, heterogeneous computing, machine learning, TICL}

\collaboration[c]{on behalf of the CMS Collaboration}

\proceeding{Calorimetry for the High Energy Frontier (CHEF2019)\\
  25--29 November 2019\\
  Fukuoka, JAPAN}

\begin{document}
\maketitle
\flushbottom

\section{Introduction}
\label{sec:intro}

The High Luminosity LHC (HL-LHC)~\cite{ApollinariG.:2017ojx} is an ambitious project that will start a new era in high energy physics, extending the present capabilities of discovering new particles and enabling the observation of rare processes that are expected to occur below the current sensitivity level. According to the current schedule ~\cite{1}, HL-LHC will achieve a peak luminosity of $7.5\times 10^{34}$ $\text{cm}^{-2}\text{s}^{-1}$, about seven times the nominal value that LHC can deliver~\cite{2}.
However, the huge amount of data provided by HL-LHC is not costless: each bunch crossing will have up to 200 simultaneous collisions (\emph{pile-up}) and the detectors will operate with higher radiation levels than in the previous runs.

\begin{figure}[htbp]
\centering 
\includegraphics[width=.4\textwidth]{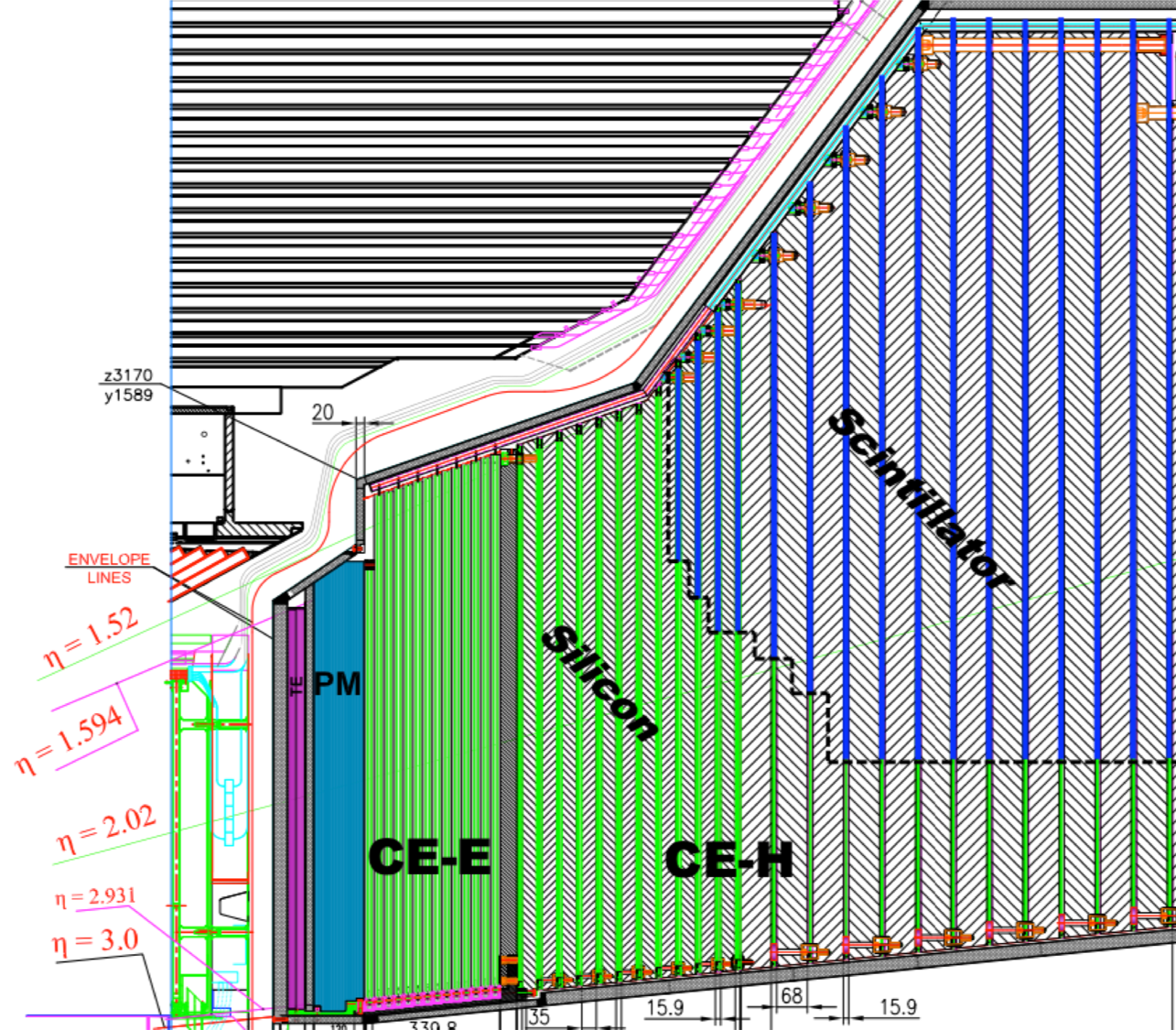}
\qquad
\includegraphics[width=.4\textwidth]{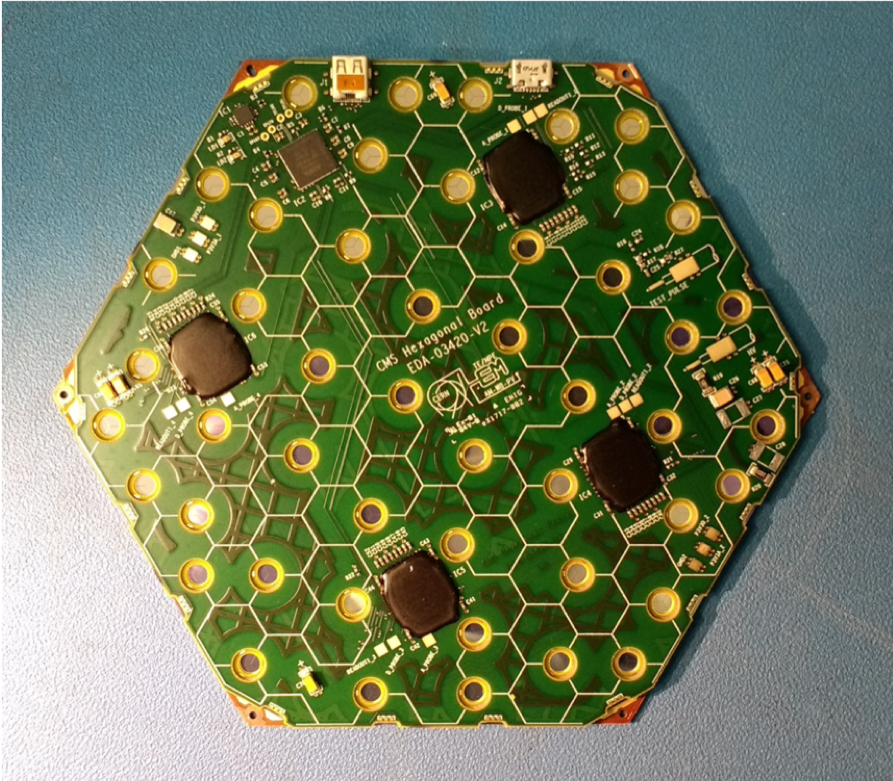}
\caption{\label{fig:hgcal} (\emph{Left}) Schematic view of HGCAL in the region $1.5 < \eta < 3.0$ and $z > 0$. (\emph{Right}) A sample hexagonal silicon sensor used in both CE-E and CE-H~\cite{3}.}
\end{figure}

To face the new challenges coming with HL-LHC, several upgrades will be applied to all major LHC experiments during shutdown periods both on the hardware and on the software sides. The High Granularity Calorimeter (HGCAL)~\cite{3} will replace the current endcap calorimeter in CMS experiment~\cite{cms} during the Long Shutdown 3 (LS3) in 2024-2026. It consists of $600$ $\text{m}^2$ of silicon sensors for a total of 6 million channels and $400$ $\text{m}^2$ of plastic scintillator tiles with SiPM readout, for a total of 240 thousand channels. The electromagnetic part of HGCAL (CE-E) consists of 28 layers per endcap, with hexagonal modules based on silicon sensors as the only active material, while Cu, CuW and Pb are used as absorber material. The radiation and interaction thicknesses are 25$X_0$ and 1.3$ \lambda$, respectively. The hadronic part (CE-H) is composed of 8 full-silicon layers and 14 Si + scintillator mixed layers, for a total of 22 layers per endcap, with stainless steel and Cu as absorber material and interaction thickness 8.2$\lambda$. The distinctive feature of this brand new detector is the hexagonal shape chosen for the silicon sensors, that provides precise measurements of particle showers. Two silicon cell sizes are forseen with an area of $\sim$0.5 or $\sim$1 $\text{cm}^2$ delivered in three different thicknesses of 120, 200 or 300 $\mu\text{m}$, while the size of scintillator tiles is 4-30 $\text{cm}^2$. Each endcap has a radius of 2.3 m and is 2 m deep, covering the region $1.5 < |\eta| < 3.0$. The weight is 215 tons per endcap and the full system operates at $-35^\circ$C. A schematic view of HGCAL and a sample of a silicon cell are shown in Figure~\ref{fig:hgcal}.

\section{Reconstruction in CMS-HGCAL}
The reconstruction software in HGCAL is being developed with speed and portability in mind. The expected CPU trend in the next years will improve software performance by a factor~$ \sim3$~\cite{4}, while offline workflows and computing in CMS will require a factor~$ \sim30$, mainly driven by the reconstruction of simulated events~\cite{5}. In order to gain the missing factor~$ \sim10$ in performance, the HGCAL software reconstruction cannot rely on any other existing sub-detector software: this new detector represents a unique opportunity to exploit modern architectures and technologies. Therefore, two main solutions are being adopted to provide the needed improvements in CMS performance for Phase-2 runs: heterogeneous computing and machine learning. On the hardware side, GPUs have shown great results in terms of speedup in recent years and the use of hybrid architectures is spreading in many fields of science. In addition, NVIDIA GPUs can be programmed with CUDA (Compute Unified Device Architecture), a parallel computing platform and programming model designed to work with programming languages such as C, C++, Fortran and Python, allowing to easily accelerate compute intensive portions of the applications~\cite{6}. On the software side, machine learning models are largely exploited to accomplish an enormous variety of tasks and can provide better results than traditional methods in many cases. Furthermore, some machine learning algorithms (e.g. Convolutional Neural Networks) can be executed on GPU, thus reducing both training and inference times, thanks to frameworks like Tensorflow~\cite{7} and Keras~\cite{8}.

\subsection{TICL: The Iterative CLustering}
The development of the HGCAL reconstruction software is driven by three main concepts:
\begin{enumerate}
\itemsep0em
    \item Particles deposit energy and create \emph{RecHits};
    \item \emph{RecHits} on each layer are clustered together to form \emph{LayerClusters} (2D objects);
    \item \emph{LayerClusters} are linked together to form \emph{Tracksters}, collections of \emph{LayerClusters}.
\end{enumerate}

\begin{figure}[tbp]
\centering 
\includegraphics[width=.9\textwidth]{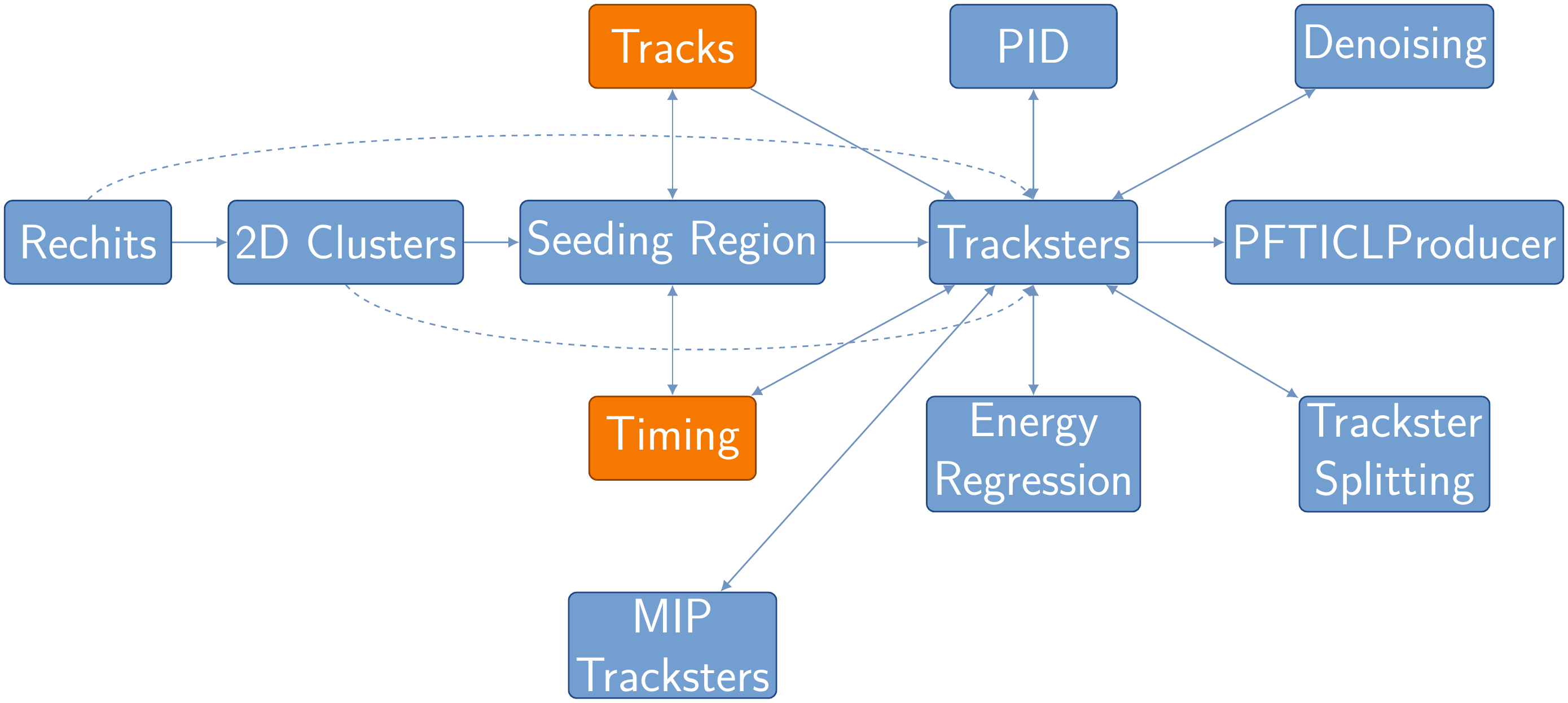}
\caption{\label{fig:ticl} Design of TICL framework. Arrows show connections between different parts of the reconstruction, pointing from the output of a step to the input of a next step. Double pointing arrows indicate a strong connection between two parts of the reconstruction. \emph{Tracks} and \emph{Timing} (orange cells) are information coming from other subdetectors (respectively Tracker and MTD)~\cite{9}.}
\end{figure}

In order to build the reconstruction chain exploiting the full potential of HGCAL, a modular framework has been developed and is constantly evolving. TICL (The Iterative CLustering)~\cite{9} modules and interfaces are defined such that new developers don't need a deep knowledge of the official CMS software core framework (CMSSW) and can easily contribute. In addition, its flexibility and modularity allow users to test their own algorithms and compare performances, as they can be plugged on top of the framework without applying any strong modification to the existing workflow. The design of TICL framework is shown in Figure~\ref{fig:ticl}. The next sections will focus on the aforementioned three main aspects of TICL reconstruction. First, a typical TICL iteration that links \emph{LayerClusters} together to produce \emph{Tracksters} will be described in Section~\ref{sec:iter}. The 2D clustering procedure with CLUE algorithm is then presented in Section~\ref{sec:clue}; finally, preliminary results of particle identification and energy regression performed on a \emph{Trackster} with a Convolutional Neural Network are shown in Section~\ref{sec:pid}.

\subsection{TICL iterations}
\label{sec:iter}
\begin{figure}[tbp]
\centering 
\includegraphics[width=.4\textwidth]{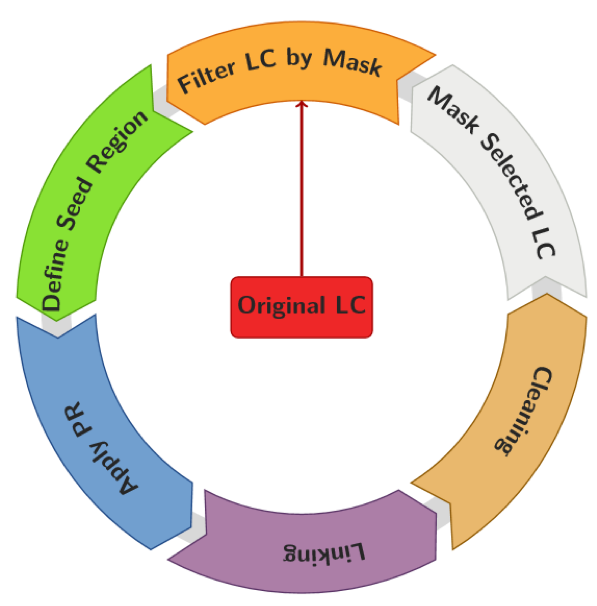}
\qquad
\includegraphics[width=.4\textwidth]{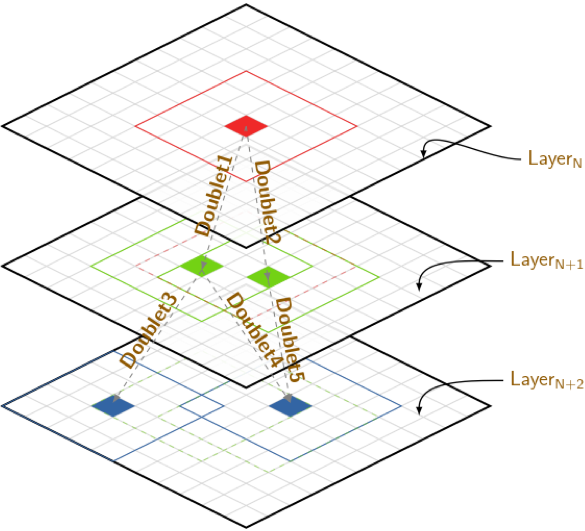}
\caption{\label{fig:iter} (\emph{Left}) Structure of a typical iteration in TICL. (\emph{Right}) Scheme of \emph{Cellular Automaton} pattern recognition implemented in TICL~\cite{9}.}
\end{figure}

\emph{Tracksters} produced by TICL iterations are suitable for representing real physics objects (i.e. particle showers). In order to guarantee such a correspondence, it's necessary to build a reconstruction workflow that takes into account the specific physics process that different particles undergo into the detector (hadronic and electromagnetic shower development, minimum ionizing particles, e.g. muons, etc.). Four different types of iteration exist now in TICL: \emph{track-seeded} (collects information from the Tracker to get the entry point and momentum direction of charged particles at the front face of the HGCAL detector), \emph{MIP} (\emph{Minimum Ionizing Particle}), \emph{electromagnetic} and \emph{hadronic}. The structure of a TICL iteration is shown in Figure~\ref{fig:iter}. A seeding region is first defined as a window in the [$\eta$, $\phi$] space on a certain layer and a pattern recognition algorithm is applied to all the available \emph{LayerClusters} within the seeding region. Then, linking, cleaning and classification tasks follow, exploiting timing information when possible. At the end of each iteration, the \emph{Tracksters} are required to pass quality criteria and particle identification: all the \emph{LayerClusters} belonging to the selected \emph{Tracksters} are masked out and are not available for the next iteration.

Thanks to the modular design of the framework, several pattern recognition algorithms can be tested to produce the best results. Currently, a \emph{Cellular Automaton} algorithm is being used based on the experience with CMS Track reconstruction~\cite{10}. It consists of six main steps:

\begin{enumerate}
\itemsep0em
    \item Start from a Layer\textsubscript{N} and consider a specific \emph{LayerCluster}
    \item Open a window in the [$\eta$, $\phi$] space around it and project it onto the next Layer\textsubscript{N+1}
    \item Consider all the \emph{LayerClusters} inside this region and try to establish a \emph{Doublet} connection between the \emph{LayerClusters} on the two adjacent layers
    \item Apply some compatibility criteria to decide if the \emph{LayerClusters} should be linked or not (i.e. geometry constraints, energy, timing compatibility, etc.)
    \item Repeat this same procedure for all the \emph{LayerClusters} on Layer\textsubscript{N}
    \item Repeat this same procedure for all pairs of contiguous layers [Layer\textsubscript{K}, Layer\textsubscript{K+1}]
\end{enumerate}

At the end of the process, pairs of \emph{LayerClusters} will be connected into \emph{Doublets}. Consecutive \emph{Doublets} (i.e. doublets that share the ``middle'' \emph{LayerCluster}) will be linked together if configurable alignment requirements are satisfied. The set of all connected \emph{Doublets} will form a direct acyclic graph that serves as building block for a \emph{Trackster}. Note that this procedure can be properly configured to allow missing consecutive \emph{LayerClusters} and establish links between non-adjacent layers (e.g. in \emph{MIP} iteration). 

\subsection{Clustering by energy}
\label{sec:clue}
Due to the high luminosity and detector granularity, each event will feature a very large number of \emph{RecHits} ($\sim10^5$). Therefore, it is important to have a clustering algorithm to collect hits together on each layer based on their energy density. The CLUE (CLUsters of Energy)~\cite{11} algorithm was designed with speed and efficiency in mind. It is GPU-friendly and constructs small clusters, reducing the number of objects in input to TICL iterations by an order of magnitude.

The first step of the CLUE algorithm consists of building a grid spatial structure to arrange \emph{RecHit} indices on each layer into tiles (whose size is much smaller than the size of a layer) that fully exploit the detector granularity and allow fast querying of $d$-neighborhood of a cell (it only requires to loop over 4 adjacent tiles at maximum, given that the distance $d$ is small). After calculating the local density for each \emph{RecHit} (step 2), its "nearest higher" \emph{nh} (index of the nearest hit with higher density) and the distance $\delta$ from it are computed (step 3). Then, seed promotion and outlier demotion is performed (step 4) with the following conditions:
\begin{align*}
\text{seeds:} \qquad \rho > \rho_c \hspace {3pt}, \hspace {3pt} \delta > \delta_s \\
\text{outliers:} \qquad \rho < \rho_c \hspace {3pt}, \hspace {3pt} \delta > \delta_o
\end{align*}
where $\rho_c$, $\delta_s$ and $\delta_o$ are tunable parameters of the algorithm. A cluster ID is assigned to all the seeds, while hits that are neither classified as seeds nor as outliers are defined as \emph{followers} of their \emph{nh} and put into their \emph{nh}'s followers list. In the final step, cluster IDs are pushed from seeds to their followers iteratively.

CLUE achieves a remarkable speedup on both CPU and GPU with respect to the previous clustering algorithm of HGCAL reconstruction. Results within CMSSW are available in~\cite{12}, while a detailed study of the performance can be found in~\cite{11}.

\subsection{Particle identification and energy regression}
\label{sec:pid}
The final purpose of the TICL framework is to reconstruct physics objects and energies and, at the same time, give probabilities on particle identification. To accomplish this task, some preliminary studies were conducted with a single particle produced in front of HGCAL in events where no pile-up is simulated. The momentum of the particle pointed from the vertex (0,0,0), center of the CMS detector. Events were simulated in such a way that particle showers (in case of electrons, photons and charged hadrons) could be fully contained inside the detector.

Since \emph{Tracksters} are suitable for representing real physics objects, a Convolutional Neural Network was designed to perform particle identification and energy regression on \emph{Tracksters} built by TICL electromagnetic iterations. In order to have fixed-sized inputs to the network, each \emph{Trackster} is represented as an image $50 \times 10 \times 3$, where the dimensions represent respectively the number of HGCAL layers per endcap, the maximum number of \emph{LayerClusters} on each layer and the number of features (energy, $\eta$, $\phi$). In this representation, each pixel of the image corresponds to a \emph{LayerCluster} that belongs to the \emph{Trackster}. Furthermore, \emph{LayerClusters} on each layer have been sorted by decreasing energy, applying a zero-padding whenever a layer featured less than 10 clusters, while removing some low energy \emph{LayerClusters} in layers with more than 10.

\begin{figure}[t]
\centering
\includegraphics[width=.4\textwidth]{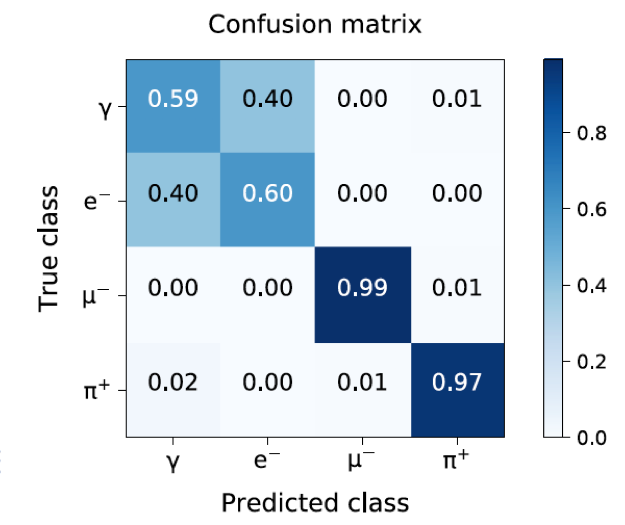}
\caption{\label{fig:pid} Confusion matrix showing the performance of particle identification.}
\end{figure}

\begin{figure}[h]
\centering
\includegraphics[width=.99\textwidth]{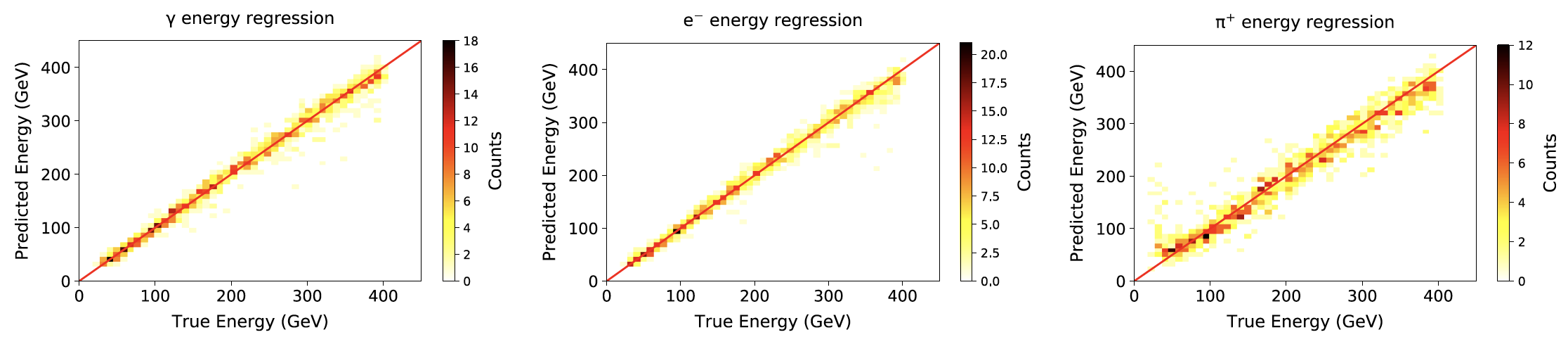}
\caption{\label{fig:er} Energy regression preliminary results for photons, electrons and charged hadrons.}
\end{figure}

A preliminary performance study was conducted on a 4-classes model (electron, photon, muon, charged hadron). The dataset consisted of 40 thousand events (10 thousand per particle type): $80\%$ has been used for training, $10\%$ for validation and the remaining $10\%$ for testing. The CNN was trained for 15 epochs (passes of the algorithm through the entire dataset), using the sum of cross-entropy~\cite{godfellow} and mean squared error~\cite{godfellow} as loss function to account for particle ID and energy regression, respectively. In order to have the value of the two functions of the same order of magnitude during training, the energies of the \emph{Tracksters} were normalized with respect to the data sample. The CNN was trained with Tensorflow~\cite{7}. Results are shown in Figure~\ref{fig:pid} and Figure~\ref{fig:er}. The confusion between electrons and photons is expected to be solved in future by exploiting information coming from the Tracker. The energy regression must be fine tuned and improvements are needed especially in the case of charged hadrons, since part of the shower is reconstructed by hadronic iteration. More classes will be added and the evolution of TICL is expected to improve the performance of this task.

\section{Conclusion}
The High Granularity Calorimeter for CMS is a very ambitious project, being a "tracking" device with high hit multiplicity and precise time information. TICL, a modular framework for particle reconstruction, aims at exploiting all the features of this novel detector. At the same time TICL is being developed with modern technologies in mind in order to improve to computing performances, both online and offline, within CMSSW in the HL-LHC era.

\end{document}